\documentclass[10pt,a4paper,oneside]{report}
\usepackage[T1]{fontenc}
\usepackage{palatino,mathpazo}
\usepackage{amsmath,amsfonts,amssymb}
\usepackage{microtype}
\usepackage{graphicx}
\usepackage[table]{xcolor}
\usepackage{hyperref}
\hypersetup{
            colorlinks = true  , 
	    urlcolor   = myBlue, 
	    linkcolor  = myRed , 
	    citecolor  = myRed   
	   }
\usepackage{wrapfig}
\usepackage[small,sf,centerlast]{caption} 
\usepackage{booktabs}
\usepackage{geometry}
\usepackage{natbib}

\usepackage[scaled=1.03]{inconsolata} 
\def\MESA{\texttt{MESA}}

\def\teff{T_{\rm eff}}

\def\rsol{R_\odot}
\def\msol{M_\odot}

\def\llsol{L/L_\odot}

\def\nabrad{\nabla_{\mathrm{rad}}}

\def\H1{^1\mathrm{H}}
\def\He3{^3\mathrm{He}}
\def\He4{^4\mathrm{He}}
\def\C12{^{12}\mathrm{C}}
\def\N14{^{14}\mathrm{N}}
\def\O16{^{16}\mathrm{O}}

\def\diff{{\mathrm d}}

\def\unity{ \hbox{1\kern-.23em l} }
\def\zero{ \hbox{0\kern-.23em |} }
\def\field{ \hbox{I\kern-.23em K} }
\newcommand{\afgtitle}[1]{{\LARGE #1}}
  
\newcommand{\afgauthor}[1]{{\large{\itshape #1}}}
\newcommand{\afgsection}[1]{{\scshape #1}}

\definecolor{Maroon}{RGB}{128,  0,  0}
\definecolor{myBlue}{RGB}{  8, 85,146}
\definecolor{myRed}{RGB} {138, 10, 11}

\pdfoutput=1
\begin{document}
 
\centerline{\afgtitle{\`{A} Propos Nonlinear Pulsations of R CrB Variables}} 

\centerline{\textcolor{Maroon}{\noindent\rule{0.93\linewidth}{0.6pt}}}

\medskip
\centerline{\afgauthor{Alfred Gautschy}}
\centerline{\textit{CBmA 4410 Liestal, Switzerland}} 

\bigskip

\noindent\makebox[\textwidth][c]{
\begin{minipage}{0.8\linewidth}
{\small \noindent Helium-star models were dynamically evolved
into the region of the HR Diagram where R~CrB variables are found. 
The \MESA\,stellar evolution code was able to pick up pulsational
instabilities with cycle lengths that are compatible with the
periods observed in pulsating R~CrB variables. The properties of the 
computed pulsations hint at their being strange modes.}  
\end{minipage}}

\bigskip\bigskip

\centerline{\afgsection{1. INTRODUCTION}}

The class of R CrB (RCB henceforth) variables is characterized by
recurring unpredictable fadings by several magnitudes. The ensuing
gradual recoveries take weeks to months. The spectra of these reddish
supergiants are peculiar because they are very hydrogen-deficient and
dominated by carbon and oxygen \citep[e.g][]{Clayton1996,
Crawford2023}. Only about 150 of these variable stars are currently
known in the Galaxy \citep{Tisserand2020}, along with a few dozen in
the Magellanic Clouds \citep{Soszynski2009}.  Hence, the evolutionary
phase that produces and maintains such objects must be short and/or
the processes making them rare.

Many RCB stars exhibit additionally cyclic low-amplitude (at most of
the order of a few tenth of a magnitude) light variability that is
attributed to radial pulsations. Although the periods cluster around
40 to 50 days, they range from about 10 to around 100 days. The
associated radial-velocity variations reach a few
km/s \citep{Lawson1997}.

\begin{figure}[h]
	\center{\includegraphics[width=0.75\textwidth]{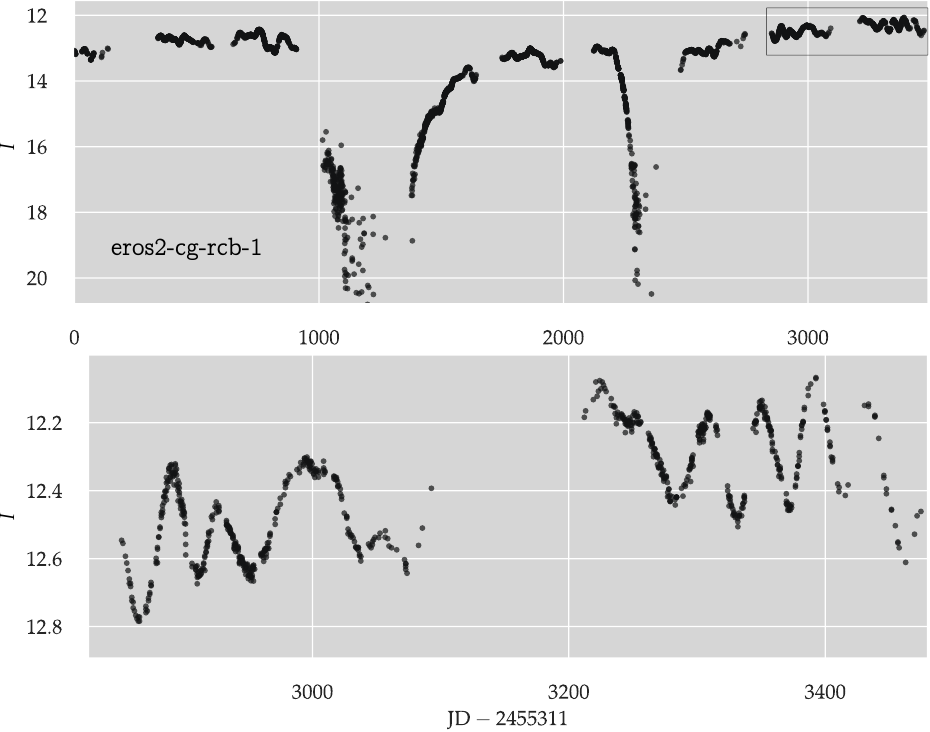}}
	\caption{An illustrative example of the pulsating RCB variables
			 as monitored in the RCOM project.  
	         The framed region on the right of the upper panel 
	         is enlarged in the lower figure to emphasize the 
	         cyclic nature of the low-amplitude photometric variability.
			}\label{fig:LightCurve_eros2}
\end{figure}

The high-quality lightcurve of a representative pulsating RCB variable, which
is monitored in the 
\href{https://ogle.astrouw.edu.pl/ogle4/rcom/rcom.html}
{\texttt{RCOM}} project \citep[cf.][]{Udalski2008} is displayed in
Fig.~\ref{fig:LightCurve_eros2}. The Galactic-Bulge RCB star
eros2-cg-rcb-1 fades dramatically every few years. Superimposed on
that, the star sports a cyclic photometric variability on the
timescale of around 40 days with an amplitude of a few tenths of a
magnitude measured in the photometric $I$-band.

The astrophysical origin of RCBs is not yet fully understood.  One
hypothesis assumes that a late He-shell flash leads to shedding the
H-rich envelope of a low-mass AGB star and the exposure of the bare,
He-dominated stellar remnant. Alternatively, hydrogen-deficient stars
might be the results of mergers of two white dwarfs of suitable
compositional flavors.  Recent stellar evolution computations sought
to identify nucleosynthesis signatures that allow to discriminate
between the conjectured formation channels
\citep[e.g.][]{Lauer2019, Crawford2020}. The computations of 
\citet{Schwab2019} showed that the evolutionary tracks of post-merger models 
and helium star models converge once they moved up to the cool
supergiant region and depart from it to evolve at essentially constant 
luminosity~--~often at $L_\ast/M_\ast \sim \mathcal{O}(10^4)$~--~to
high effective temperatures independent of their past.   
 
It was the very family of pulsating RCB stars where the concept
of \emph{strange modes }emerged \citep{Wood1976,Cox1980}.  The linear
non-adiabatic pulsation modeling of RCB stars led to pulsation modes
that have no counterpart in the adiabatic limit of canonical sound
waves. Furthermore, strange-mode instabilities grow on very short
timescales \citep[e.g.][]{SBG98, Saio2008}.
Recently, \citet{Gautschy2023} stumbled over a pulsational instability
of early~--~post-AGB star models that is reminiscent of strange
modes. Because RCB stars live in the same neighborhood on the HR plane
as the low-mass stars at the end of their AGB evolution, 
a check whether the same kind of pulsational instability can be
picked up and followed with \MESA\,in RCB stars too is an obvious step to take.

This \emph{engineering }report presents exploratory dynamical 
evolution computations with \MESA\,on helium star models (Sect.~2), 
the pulsations that were picked up along the way (Sect.~3), and 
finally points out remaining problems and emphasizes 
the potential of this approach (Sect.~4). 

\bigskip
\centerline{\afgsection{2. EVOLUTION COMPUTATIONS}}

\begin{wrapfigure}{l}{0.40\textwidth}
	  \vspace{-10pt}
	  \begin{center}{
			\includegraphics[width=0.40\textwidth]{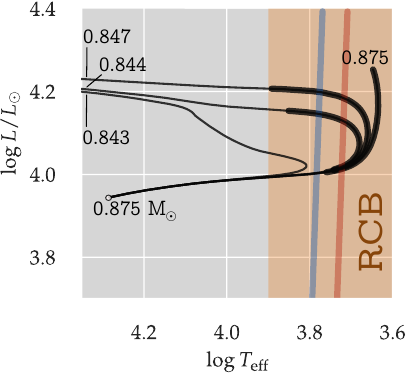}	
	  }\end{center}
	  \vspace{-15pt}
	  \caption{Evolutionary tracks of He-star models on the HR plane. 
			 RCB variables are observed in the highlighted region.
	  	      }\label{fig:EvolutionHeStars}
	 \vspace{-25pt}	
\end{wrapfigure}

Stellar models referred to henceforth were computed with 
\MESA\texttt{star}~in the version close to what
is described in \citet{Paxton2019}. The computational setup is based
on what is provided in the \texttt{test\_suite/R\_CrB\_star} case of
\MESA\,version r$21.12.1$. The start model is a $0.875\,\msol$ helium star
in advanced He-shell burning (open circle in Fig.~\ref{fig:EvolutionHeStars}) 
that evolved from the HeZAMS with 
$Y_{\mathrm{initial}} = 0.994, Z_{\mathrm{initial}} = 0.006$. 
The details of the evolution computations that lead to the start-model 
provided by the \texttt{test\_suite} case \texttt{R\_CrB\_star} can be found 
in \citet{Schwab2019}.

The stellar-physical calibration of the observed RCB variables remains
uncertain even in the era of GAIA. Photometric and spectroscopic
calibrations frequently deviate from each other so that
Fig.~\ref{fig:EvolutionHeStars} shows only a rectangle that
hints at the region where RCB variables are typically observed.

The \MESA-computed RCB-like models develop lumi\-nosi\-ty-to-mass ratios that 
exceed $10^4$ once they evolve through their turn-around as cool 
supergiants on the HR plane. Most importantly, the models 
evolve into and through the orange area in Fig.~\ref{fig:EvolutionHeStars}
that encloses the region where RCB variables are observed; 
this makes the evolutionary models interesting enough for a closer 
look at their pulsational stability behavior.

For the purposes of this report, the above mentioned start model
was evolved in dynamical mode: The acceleration term in 
the momentum equation was switched on by setting 
\texttt{change\_v\_flag = .true.} and \texttt{new\_v\_flag = .true.} 
in the \texttt{inlist} file. The model star's evolution was followed 
as far as possible through the region on the HR plane where RCB stars 
are found in nature. In the \emph{temporal coarse-resolution }runs, 
the canonical evolutionary timesteps 
varied from hundreds to a few year at the end. For \emph{temporal
high-resolution} studies, model snapshots suitable for \MESA\,restarts 
(referred to as \texttt{photos}) were frequently stored.  

The locus of the  conservative evolution of the $0.875\,\msol$ terminated 
in the orange region of Fig.~\ref{fig:EvolutionHeStars} with convergence
problems (see Sec.~3, discussion of epoch D).
The classical instability strip (a blue and a red line serve as rough
references of the respective edges) was crossed and the model star 
rose its luminosity on the verge to turn around to higher 
effective temperatures. Before successfully doing so, 
the code ran into convergence problems 
\citep[see also][on similar experience]{Schwab2019}. Eventually, the
computations stalled with exceedingly small timesteps.

To probe additional regions on the HR plane populated by RCB stars,
it seemed easiest to evolve the $0.875\,\msol$ He-star starting model 
by invoking various intensities of mass loss: 
Bl\"ocker-type mass loss, which is implemented in 
\MESA\,\citep[cf.][]{Paxton2011}, was activated and tuned with the parameter
\texttt{Blocker\_scaling\_factor}. Figure~\ref{fig:EvolutionHeStars} 
displays three additional evolutionary sequences that were obtained for the choices
\texttt{Blocker\_scaling\_factor}$\,\equiv \eta_{\mathrm{B}} = 0.0005,\,0.001,$ 
and $0.005$. The larger the scaling factor, the higher is the mass-loss rate. Hence, 
the resulting stellar masses at the end of the respective tracks 
identify therefore uniquely the applied scaling factors. 
Even though the masses of the different tracks 
differ by less than $5\%$, they probe a disproportionately large area of 
the RCB region. The particular choices of the numerical values of 
\texttt{Blocker\_scaling\_factor} were not physically motivated. 
The requirement was nothing more than taking a model star through 
the RCB region on the HR plane with a total mass that does not contradict 
the currently adopted calibrations of the stellar parameters of the 
RCB variables. 
\emph{No pulsations developed at any time in the evolutionary calculations 
	  with coarse temporal resolution.}
 
To find potential pulsations, evolution is not allowed to proceed at
computational timesteps that exceed the expected pulsation periods. 
Therefore, the collection of \texttt{photos} models served as starting epochs 
for temporal high-resolution computations. The computational time steps 
were capped at about $1$~day. Additionally, many initial models were  
numerically perturbed. The easiest way was to request a
change of the spatial resolution of the initial models: In comparison to the lower 
time-resolution runs, the temporal high-resolution computations were started  
requesting a spatially denser mesh with 
\texttt{mesh\_delta\_coeff < 1} (for the results presented here,
a value of 0.7 was adopted. In the range 0.7 - 0.9, the particular choice 
did not matter; smaller values are computationally more expensive though).
This leads locally to the insertion of additional computational cells by 
interpolation; the resulting small numerical noise turned out 
to be sufficient to get pulsations started in appropriate   
model stars.\footnote{However, the same procedure failed for 
Cepheid-like $7.5\,\msol$ model stars in the classical instability strip.} 
With this prescription, pulsations could indeed be picked up with \MESA. 

In Fig.~\ref{fig:EvolutionHeStars}, the pulsationally unstable epochs 
along the evolutionary tracks are connected by fat lines . 
All the three tracks that also cross the
classical instability strip developed pulsations on their way to the 
high-luminosity branch in the yellow/red supergiant region. 
Only the models along the lowest-mass track remained pulsationally 
stable at any time. 

Most of the high-resolution runs were done with the default outer
boundary condition (OBC) that computes the appropriate surface
pressure (via calling routine \texttt{set\_Psurf\_BC}). Additional OBC
options were tried out:
\texttt{use\_momentum\_outer\_BC} and \texttt{use\_compression\_outer\_BC}
\citep{Paxton2015}. Regardless of the choice, the character of the results persisted. 
Most importantly, the models could not be stabilized to the point
where more of them settle into limit cycles. In most cases, the default
OBC and the \texttt{use\_momentum\_outer\_BC} yielded essentially the
same results. Requiring \texttt{use\_compression\_outer\_BC}
destabilized the models additionally so
that \texttt{mesh\_delta\_coeff} smaller than unity led to weaker convergence
during the start-up phase.

To contrast the instability domain hinted at by the nonlinear
computations of this report with linear pulsation theory: The HR plane 
clipped in Fig.~\ref{fig:EvolutionHeStars} lies completely inside the region 
of radial pulsation instabilities obtained from nonadiabatic stability
analyses reported in \citet{Saio2008}.

\bigskip
\centerline{\afgsection{3. PULSATION PHENOMENOLOGY}}

\begin{wrapfigure}{r}{0.37\textwidth}
	  \vspace{-20pt}
	  \begin{center}{
			\includegraphics[width=0.37\textwidth]{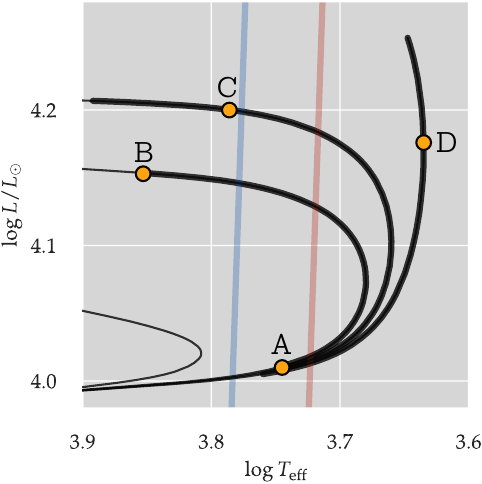}	
	  }\end{center}
	  \vspace{-15pt}
	  \caption{Epochs A to D whose pulsational characters are set out
			 in the text.
	  	      }\label{fig:PulsationsAtoD}
	 \vspace{-34pt}	
\end{wrapfigure}
Characteristic pulsation properties that were encountered in this 
exploratory exercise are highlighted at selected epochs, labeled A to D in
Fig.~\ref{fig:PulsationsAtoD}. 

\textsc{Epoch A, }on the lower-luminosity branch of the helium-star models 
that evolve into the RCB region, on the HR plane is in the
neighborhood of the onset of pulsations. It is representative for all
sequences that eventually developed pulsations. This is not surprising
because the respective evolutionary tracks do not yet diverge much and
the difference in stellar mass is minuscule.

\begin{figure}[h!]
  \centering
  \includegraphics[width=0.49\textwidth]{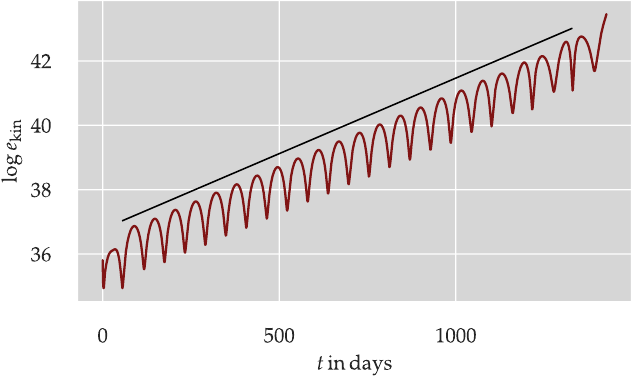}
  \hfill
  \includegraphics[width=0.49\textwidth]{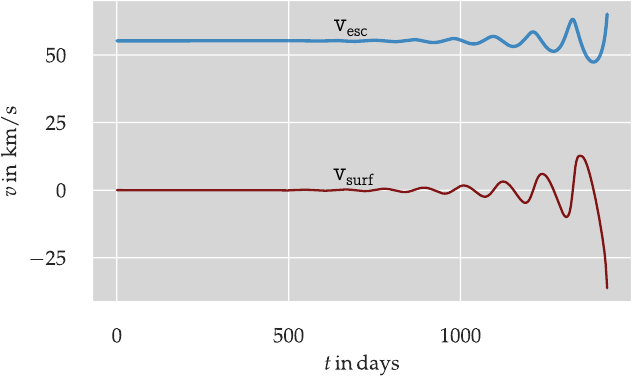}
  \caption{Epoch A: The pulsation dynamics encountered during the onset
  of the instability. The time in days along the abscissa is measured 
  relative to the starting epoch in the plot.}\label{fig:CaseA_dynamics}
\end{figure}

Already very early on in the instability domain, the exponential
growth rates are encountered.  For the model whose
behavior is plotted in Fig.~\ref{fig:CaseA_dynamics},
at $(\log \teff,\,\log \llsol)$ = (3.745, 4.010) and a stellar mass of
$0.8737\,\msol$ ($\eta_{\mathrm{B}} = 0.001$), the kinetic energy
(measured in erg) of the star grows by six orders of magnitude over
twelve pulsation cycles.  The average cycle length
($\Pi_{\mathrm{puls}}$) is about 115~days (the time for two bumps in
the left panel of Fig.~\ref{fig:CaseA_dynamics} or peak to peak in the
surface velocity in the right-hand panel).  The quantity
$\eta \doteq \Pi_{\mathrm{puls}}/\tau_{\mathrm{e-folding}}$ measures
the growth rate of the instability. The e-folding time of the
amplitude growth, $\tau_{\mathrm{e-folding}}$, is obtained from the
slope $m$ of the fit to the $\ln e_{\mathrm{kin}}$ curve (like the
thin black line in the left panel of Fig.~\ref{fig:CaseA_dynamics}):
$\tau_{\mathrm{e-folding}} = 1/m \approx 93\,$d.  Hence: $\eta \approx
1.2$.

The way the surface pulsation velocity grows and how it compares to the
escape speed from the surface is shown in the right panel of
Fig.~\ref{fig:CaseA_dynamics}. Within six pulsation cycles, the surface 
radial velocity grows to a few~km/s. The code stalls when the model star
contracts at about $30$~km/s. This is well below the escape speed of about
$60$~km/s. The convergence problems of \MESA\, appear to be associated with 
the occurrence of a very pronounced opacity spike induced by the first partial
ionization of helium in the vicinity of the photosphere.

The pulsation that develops at epoch~A, close to the blue instability
edge along the low-luminosity branch of the helium-star evolutionary track,
grows at a rate that far exceeds that of classical pulsators. The computations
stall during the exponential growth; even above a total kinetic energy
of $10^{42}$~erg, no signs of saturation developed. Even though
the pulsation appears to be a very dynamic phenomenon, the (kinetic) energy
associated with it is very small compared to the 
$\approx -5\times 10^{49}$~erg of total energy stored in the star.

\textsc{Epoch B} illustrates that 
not all pulsations picked up by \MESA\, grow beyond physical bounds or the
numerical capabilities of the evolution code. Limit cycles were encountered
close to the blue instability edges on the high-luminosity branches
of the evolutionary tracks computed with $\eta_{\mathrm{B}} = 0.0005$ and
$0.001$, respectively. For illustration, pulsation analyses that rely 
on the cyclicity of the phenomenon are presented for Blöcker 
$\eta_{\mathrm{B}} = 0.001$  case at 
$(\log \teff,\,\log \llsol)$ = (3.853, 4.153) with $0.8458\,\msol$ at the
onset of the pulsation.

\begin{figure}[h!]
  \centering
  \includegraphics[width=0.49\textwidth]{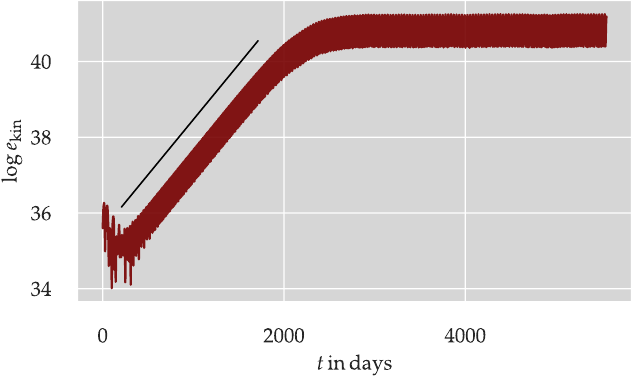}
  \hfill
  \includegraphics[width=0.49\textwidth]{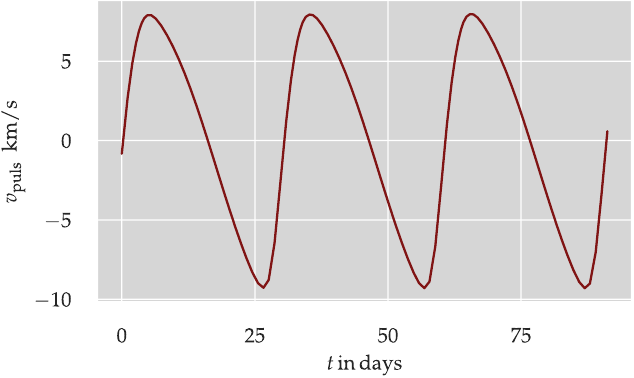}
  \caption{Epoch B: Pulsation dynamics (left) and limit-cycle pulsation velocity
  at the star's surface (right).}\label{fig:CaseB_dynamics}
\end{figure}
                 
At epoch B,  the kinetic energy of the pulsating star grows by 6 orders of magnitude
to about $10^{41} $erg where it saturates. The average cycle length is 
$\Pi_{\mathrm{puls}} \approx 30$~days; $\Pi_{\mathrm{puls}}$ varies by a   
few days around the mean even late in the saturated phase. 
The black line, fitting the envelope of the growth of the kinetic energy,
in the left panel of Fig.~\ref{fig:CaseB_dynamics} gives an 
e-folding time, $\tau_{\mathrm{e-folding}} \approx 146\,$d, yielding hence 
$\eta \approx 0.2$. 

To gain some insight into the limit-cycle variability of this model, three cycles
are picked out towards the end of the integration span. The surface 
pulsation-velocity is shown in the right panel of 
Fig.~\ref{fig:CaseB_dynamics}. To keep the time coordinate simple, 
the plotted quantity $t$ is measured relative to the starting 
model-age of the three-cycle zoom-in. The pulsation velocity is slightly 
asymmetric in amplitude and form: It changes between about -9 km/s and + 8 km/s; 
the maximum expansion speed is far from the escape speed at around 65 km/s. 

\begin{figure}[ht!]
  \centering
  \includegraphics[width=0.60\textwidth]{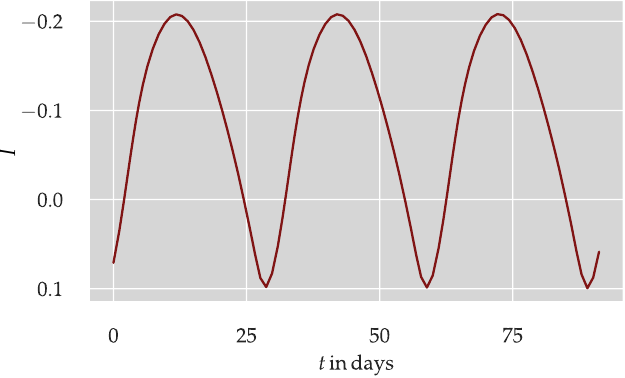}
  \hfill
  \includegraphics[width=0.36\textwidth]{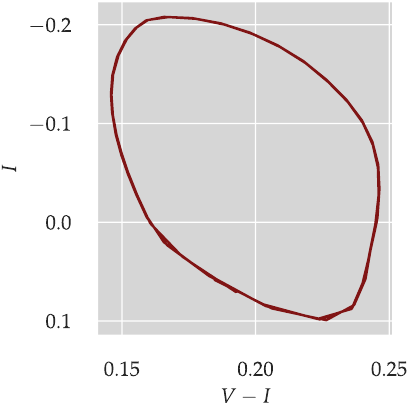}
  \caption{Epoch B: Photometric variability in the
           Johnson-Cousins system, with arbitrary zero-points
           of the magnitudes.}
          \label{fig:CaseB_photometry}
\end{figure}

The photometric signature of the cyclic variability at epoch~B is shown 
in Fig.~\ref{fig:CaseB_photometry} for the Johnson~--~Cousins passbands $V$ and $I$.

The shape of the $I$-band lightcurve is smooth and triangular with broader 
maximum light and comparatively sharp minima 
(left panel of Fig.~\ref{fig:CaseB_photometry}). 
The amplitude reaches 0.3 magnitudes. The loops traced 
out by the pulsating model star on the color-magnitude diagram, $I$ vs. 
$V-I$, are very pronounced but oriented as customary for pulsating stars. 
Classical pulsators usually sport shallower loops though.  

\begin{wrapfigure}{l}{0.5\textwidth}
	  \vspace{-26pt}
	  \begin{center}{
			\includegraphics[width=0.5\textwidth]{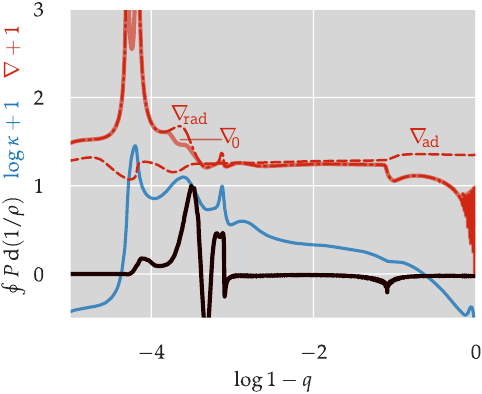}	
	  }\end{center}
	  \vspace{-15pt}
	  \caption{Epoch B: Work integral over the central cycle of the
	selection shown in Fig.~\ref{fig:CaseB_photometry} together
	with useful physical quantities.
	  	      }\label{fig:CaseB_WorkIntegral}
	 \vspace{-20pt}	
\end{wrapfigure}

The strict periodic pulsations that develop at epoch B allow 
for integrations over a cycle to learn about work done on the pulsation 
throughout the star. The differential work integral 
\[
\Delta W =       \oint\displaylimits_{\mathrm{cylce}}\! P\,\diff_t \Delta V \diff t
         \propto \oint\displaylimits_{\mathrm{cylce}}\! P\,\diff 
                                              \left(\frac{1}{\rho}\right)\,,
\]
per mass shell $\Delta m$ over a pulsation cycle informs about which
regions feed the pulsation ($\Delta W > 0$), which ones damp it
($\Delta W < 0$), and eventually, which ones are indifferent to it
($\Delta W = 0$).  The full black line in
Fig.~\ref{fig:CaseB_WorkIntegral} traces the work integral,
normalized to maximum driving, as a function of a stretched
relative-mass ($q$) coordinate. The convective structure of the star
can be deduced from the gradients $\nabla = \diff \log T / \diff \log
P$, the adiabatic, the radiative, and the effectively prevailing one ($\nabla_0$)
as red lines in Fig.~\ref{fig:CaseB_WorkIntegral}.  The spatial
variation of the Rosseland opacity is plotted (vertically shifted)
with the blue line. The three driving peaks in $\Delta W $ are
associated with the three pronounced opacity peaks.

The driving regions are mostly convective. The 
respective convection zones are, however, very nonadiabatic 
($\nabla_0 \approx \nabrad$) so that the energy flux transported by 
convection is small compared to the radiative flux. 
Therefore, the effect of convective damping/driving is likely small.

The $\Delta W $ depression very deep in the stellar interior (around
$\log\,(1-q) \approx -1$ in Fig.~\ref{fig:CaseB_WorkIntegral}) 
is induced by the nuclear-burning shell. Additionally, non-vanishing, 
although small dissipation prevails throughout most of the stellar interior. 
This a behavior is usually not accessible to nonlinear pulsation computations 
because such simulations are constrained to the low-temperature, 
nuclear-burning~--~free parts of the stellar envelopes. Here, in contrast,
the nuclear network is computed at every timestep. 
 
\begin{figure} [h!]
  \centering
  \includegraphics[width=0.50\textwidth]{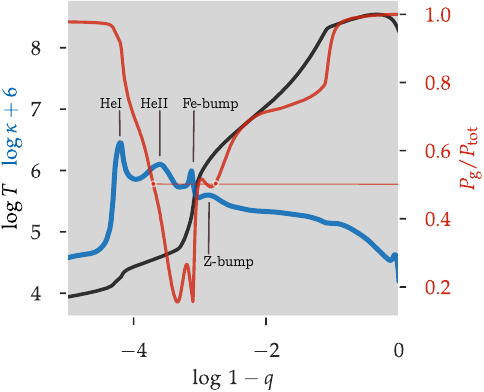}
  \hfill
  \includegraphics[width=0.4\textwidth]{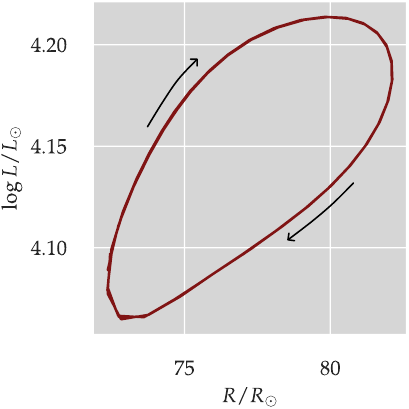}
  \caption{Epoch B: Pulsation physics (left) and visualization of the unusual
  phase lag (right).}\label{fig:CaseB_PulsPhys}
\end{figure}
The left panel of Fig.~\ref{fig:CaseB_PulsPhys} helps to physically pin down
pertinent features of Fig.~\ref{fig:CaseB_WorkIntegral}.
The black line traces the temperature profile as a function of mass
depth.  Together with the molecular-weight profile (not shown) the
three dominant opacity peaks can be attributed to (in direction of
decreasing mass) HeI and HeII partial ionization, and to the Fe-bump
caused mostly by many intra-shell bound~--~bound transitions. The
Z-bump, also due to bound-bound transitions of additional heavy
elements, slightly above $10^6$~K is mentioned for completeness: it
does not contribute to the pulsation energetics. Most of the pulsation
driving comes from the partial second ionization of He, supported by
the Fe-bump.  The red profile quantifies the importance of radiation
pressure throughout the star. The red horizontal line helps to identify
the region where radiation pressure dominates.  Most of the pulsation 
driving originates hence from where radiation pressure contributes
80\% or more to the total pressure.

The right-hand panel of Fig.~\ref{fig:CaseB_PulsPhys} emphasizes
the peculiar character of the pulsations computed here on the 
radius~--~luminosity plane. Evidently, the phase of minimum radius
coincides roughly with minimum luminosity. The phase agreement is
the same for maximum radius and luminosity. Hence, the phase lag
\citep[e.g.][]{jpc80} that measures the phase difference between
maximum light and minimum radius is about $0.5\,\Pi_{\mathrm{puls}}$
in the pulsating RCB models. In contrast, classical pulsators 
usually sport a phase lag of $0.1 - 0.2\,\Pi_{\mathrm{puls}}$.

\begin{figure}[h!]
  \centering
  \includegraphics[width=0.47\textwidth]{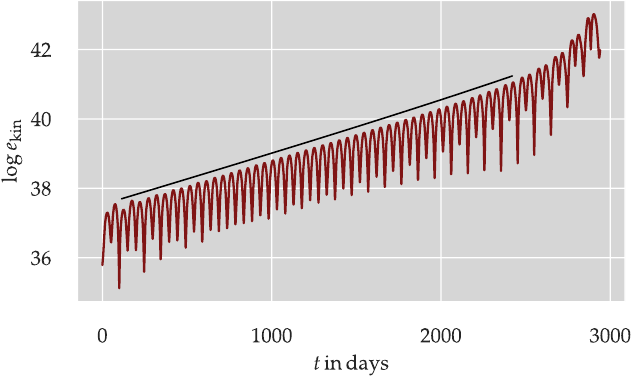}
  \hfill
  \includegraphics[width=0.47\textwidth]{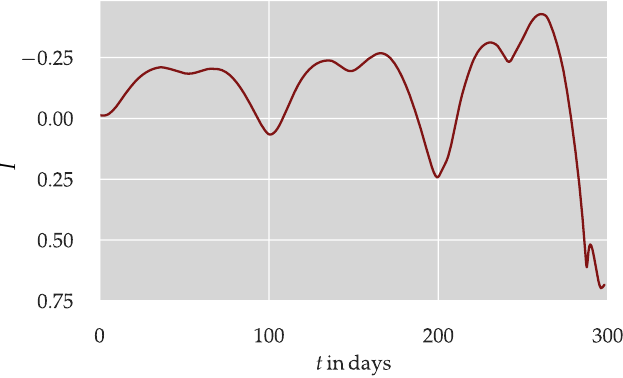}
  \caption{Epoch C: Complex nonlinear phenomenology.}\label{fig:CaseC_PulsPhys}
\end{figure}

\textsc{Epoch C} is an example of the development of a complex nonlinear 
behavior of the pulsation. In this case, no limit-cycle develops and 
the code stalls once the amplitude grows exceedingly. 
The temporal high-resolution sequence starts at 
$(\log \teff,\,\log \llsol)$ = (3.786, 4.20) with a model of 
$0.849\,\msol$ that results from the $\eta_{\mathrm{B}} = 0.0005$ evolutionary
sequence.

The kinetic energy of the star grows by five orders of magnitude
before the simulation stalls (left panel of
Fig.~\ref{fig:CaseC_PulsPhys}). The slope of the fitted line to the
growing kinetic energy curve at early times gives $\tau_{\mathrm{e-folding}} =
1/m \approx 229\,$days.  During the last few cycles, the
kinetic-energy growth seems to go even super-exponential. As usual
in such a stagnation, the timestep drops because the code tries to
resolve, in space and time, a very sharp opacity peak close to the
photosphere. The time between successive deep minima of the lightcurve
results in an average $\Pi_{\mathrm{puls}} \approx 98$~days. Once the 
dynamics becomes sufficiently nonlinear and the photometric $I$-band amplitude 
reaches a few tenths of a magnitude, double-peaked maxima develop (right panel of
Fig.~\ref{fig:CaseC_PulsPhys}). The cycle-length grows to about
100~days once the star goes terminally nonlinear.  Even under such
a large variation of the pulsation energetics, the cycle-length
responds only weakly.  Hence, this model achieves again a large value
for $\eta:\, 98/229 \approx 0.4$.
 
\begin{figure}
	\center{\includegraphics[width=0.95\textwidth]{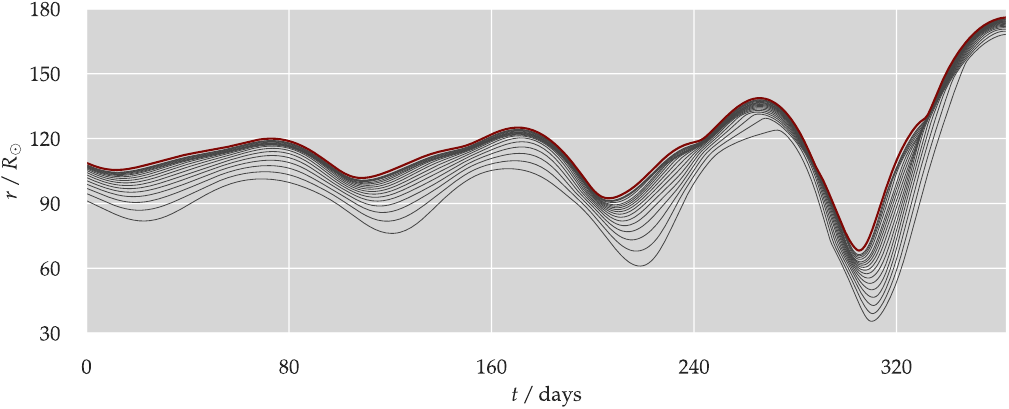}}
	\caption{Epoch C: Radius variation of selected mass shells over the 
	         last few cycles that could be computed
	         .
			}\label{fig:CaseC_RadiusShells}
\end{figure}

The double-peaked light variability of the model can be traced back to
the respective radius variation of different layers across a pulsation
cycle.  Figure~\ref{fig:CaseC_RadiusShells} illustrates, for the time
window used in Fig.~\ref{fig:CaseC_PulsPhys}, the temporal radius
variation of selected mass shells: The bottom line traces the radial
motion at $\log 1-q = -3.0$. The next higher-up lines are separated by
$\Delta \left(\log 1-q\right) = -0.05$. Eventually, the red line traces the motion
of the photosphere.
\begin{wrapfigure}{r}{0.4\textwidth}
	  \vspace{-18pt}
	  \begin{center}{
			\includegraphics[width=0.4\textwidth]{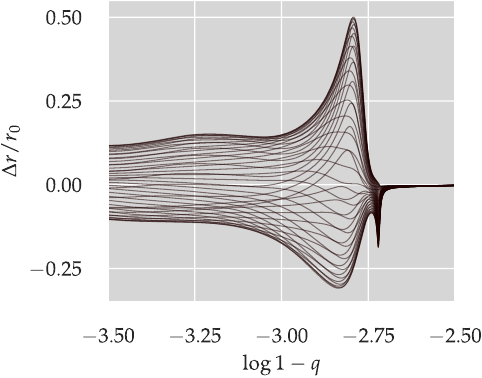}	
	  }\end{center}
	  \vspace{-15pt}
	  \caption{Epoch C: $\Delta r / r_0$ profiles over 
	  the second cycle shown in Fig.\ref{fig:CaseC_RadiusShells}.
	  	      }\label{fig:CaseC_dR_o_R0_profiles}
	 \vspace{-15pt}	
\end{wrapfigure}
The last two cycles shown in
Fig.~\ref{fig:CaseC_RadiusShells} point out that the double peaks in
the lightcurves are caused by a strong cyclic traveling wave that
comes from the deeper interior and overtakes the smaller-amplitude
radial oscillation in the most superficial layers.
The interaction of the two movements generates a compression and eventually 
a shock-front that travels outwards during the expansion phase. The shock arrives
at the photosphere before maximum radius. Subsequently, a compression feature
starts to propagate inward during early light decline
(cf. Fig~\ref{fig:CaseB_PulsPhys}), the phase which is frequently
found to be the source of convergence problems and continued
time-step reductions to a standstill.

Plotting a collection of relative radius-variation profiles at
numerous phases of a pulsation cycle visualizes which regions of the
star participate in the pulsation and how the amplitudes of the
motion vary throughout the envelope.  The mass coordinate of the abscissa
is representative for all models studied here.  Evidently, the pulsations
are efficiently quenched at around $\log 1-q = -2.7$.  In
Fig.~\ref{fig:CaseC_dR_o_R0_profiles}, the abscissa is clipped at
$\log 1-q = -2.5$ because the relative displacement profiles have no
noticeable amplitude deeper in the star. The largest displacements are found at 
around the excitation regions of HeII partial ionization and the Fe opacity bump. 
Maximum expansion and compression are not symmetric. The maximum relative
expansion in the driving region reaches 50\%. On the other hand, the
star compresses only by about 25\% at the same mass depth. In the
outermost $10^{-3}$ of the stellar mass, the radius variation flattens
out spatially.  All the essential pulsation dynamics is confined to
the just mentioned radiation dominated driving region and this dynamics 
has a pronounced running-wave character.
   
\begin{figure}
  \centering
  \includegraphics[width=0.47\textwidth]{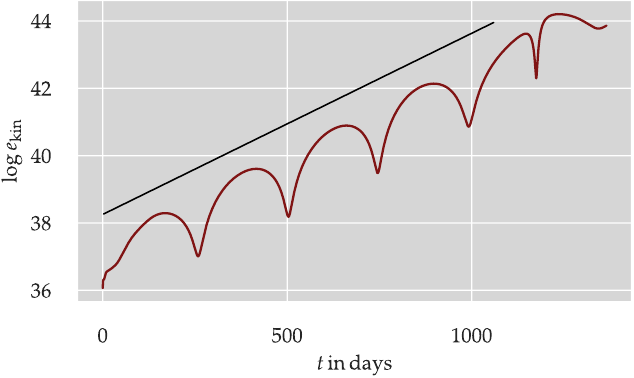}
  \hfill
  \includegraphics[width=0.47\textwidth]{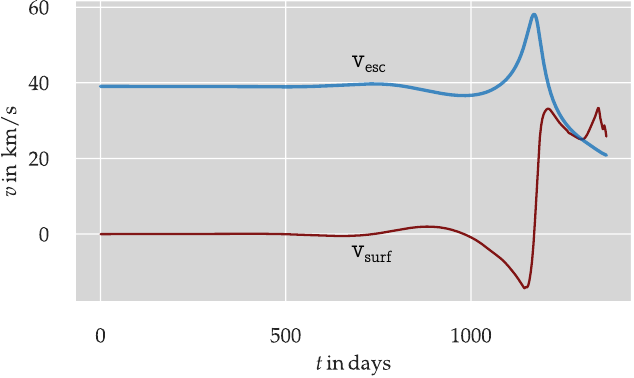}
  \caption{Epoch D: The most rapid growth of the pulsational instability at
  very cool stages.}\label{fig:CaseD_StrongInstability}
\end{figure}

\begin{wrapfigure}{l}{0.4\textwidth}
	  \vspace{-23pt}
	  \begin{center}{
			\includegraphics[width=0.4\textwidth]{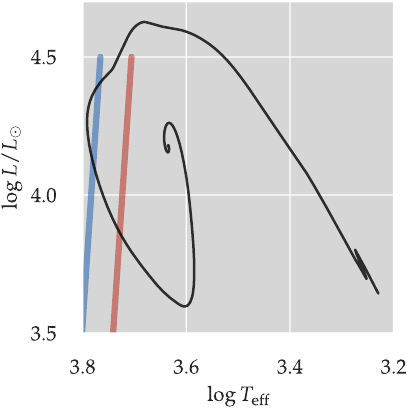}	
	  }\end{center}
	  \vspace{-15pt}
	  \caption{Epoch D: Locus traced out on the HR plane within a few
			 years, once the pulsation gets going.
	  	      }\label{fig:CaseD_HRDtrack}
	 \vspace{-23pt}	
\end{wrapfigure}
\textsc{Epoch D }illustrates the dynamics of the most unstable model stars, 
mostly they are very cool ones. Suppressing the dynamical degree of freedom
in canonical evolution computations is likely the cause of the convergence
problems that can or cannot be ironed over. 
The behavior of the model from the conservative $0.875\,\msol$ track  
at epoch D, at $(\log \teff ,\, \log \llsol)$ = (3.635, 4.176) is also reminiscent 
of what happens at the very \emph{end }of the respective computable evolutionary 
track~--~just the cycle lengths differ.

Over the roughly three cycles that were computationally accessible and
depicted in the left panel of Fig.~\ref{fig:CaseD_StrongInstability}, the
kinetic energy grows roughly by eight orders of magnitude
(i.e. $\tau_{\mathrm{e-folding}} = 1/m \approx 80\,$d).  Accordingly,
under such violent conditions, the pulsation velocity at the stellar
surface exceeds escape speed quickly, i.e. after the third and last cycle (right
panel of Fig.~\ref{fig:CaseD_StrongInstability}). Even though
determining a cycle length strains the scope of this concept, two full cycles
suggest $\Pi_{\mathrm{puls}} \approx 500$~days. Hence, $\eta = 500/80 \approx
6.3$.

The rapid growth of the model's pulsational instability makes it cover
a large area on the HR plane (Fig.~\ref{fig:CaseD_HRDtrack}). As in
most cases with unlimited growth, the code stalls during the
luminosity decline when very a sharp opacity feature develops in the
most superficial layers. The stellar radius increases from initially
about $200\,\rsol$ to finally over $700\,\rsol$. This all happens
within about four years. 
\newpage

\bigskip
\centerline{\afgsection{5. DISCUSSION AND CONCLUSIONS}}

Pulsational instabilities have been computed with stellar-evolution
codes in the past, \citep[e.g.][]{Yoon2010}. Also \MESA\, proved to be
capable of doing so \citep[e.g.][]{Paxton2013,Gautschy2023}. 
This report adds another example, with \MESA\,
being able to pick up radial pulsations, namely for RCB-like helium stars. 
In the RCB instability region, a small domain was encountered where the nonlinear
pulsations saturated and approached a limit-cycle~--~like behavior.
Very few pulsation evolved into saturation, though. 
A source of energy loss~--~possibly implementable via a proper OBC~--~is missing 
in the current setup of the computations. The presently implemented OBCs 
in \MESA\,all lead to the same long-term behavior, mostly without saturation
of the pulsations. 

The inflated structure of RCB stars induces a rather smooth transition
from the stellar surface to the circumstellar neighborhood.  Wave
reflection close to the surface is weak, as can be seen, for example,
in the plots of characteristic critical frequencies of RCB like model
stars in \citep{Wong2023}. The calculations discussed here are too
sterile to do justice to the nature of the complex circumstellar
environment of RCB stars, which likely feeds back on the
pulsations. On the other hand, the pulsations themselves might serve as
momentum and energy sources that influence dynamical dust formation
and destruction around the star. Therefore, coupling a \MESA-type
pulsation computation with some form of dynamic dust-formation
prescription has promise to answer the old question if the irregularly
occurring RCB fadings are affected by RCB-type pulsations.

The exceedingly large growth rates $\eta$ encountered in the pulsations
reported here~--~$\eta_{\mathrm RCB} = \mathcal{O}(1)$, whereas  
$\eta \lesssim \mathcal{O}(10^{-3})$ for classical pulsators~--~and the 
strong confinement of the pulsation-amplitudes
to the radiation-dominated layers of stars with $L/M \gtrsim 10^{4}$ suggest
that the pulsations are of strange-mode type.

The cycle lengths of the computed nonlinear pulsations and the light- and
pulsation-velocity amplitudes of the few examples that transition to a
limit cycle behavior indicate that strange-mode pulsations can indeed be
responsible for the cyclic variability of at least some of the RCB
stars. The instability domain hinted at here is smaller than what
observations of pulsating RCB stars require. In particular, \MESA\,did
no reveal pulsations for $\log\teff > 3.9$ at high
luminosities. Pulsating variables in the region connecting RCB stars
and eHe stars \citep[cf.][]{Saio2008} cannot be explained with 
pulsation modes as found here. Some local friction
in the \MESA\,numerics might still affect the results. 
This is also supported by
the inability to get Cepheid pulsations going with the present recipe.
On the other hand, also regular \emph{p}-mode pulsations are likely
involved in the plethora of cyclically variable hydrogen-deficient
stars \citep{Wong2023}.

Because the pulsations of \emph{full }stellar models were computed
with a dynamical stellar-evolution code, including a nuclear-burning
network, the reaction of the helium-burning shell on the pulsation can
be studied.  The work integral (Fig.~\ref{fig:CaseB_WorkIntegral})
shows that the nuclear-burning shell has a damping effect; its overall
contribution is not important, but noticeable
nonetheless. Furthermore, even the deep interior of the star maintains
a small damping effect on the stars' pulsations, this despite the very
small amplitudes of the quenched pulsation modes.

In the pulsation-driving regions matter is convectively unstable. 
With the current \MESA\,ansatz, convection was instantaneously adapting.  
The effects of \emph{time-dependent }convection on the driving/damping 
still need to be clarified. However, the pulsations are limited to 
low-density, radiation-dominated regions where convection contributes only
little to the energy transport and might therefore 
also be irrelevant for the overall energetics of the pulsations.

If DY~Per variables \citep[e.g.][]{Nikzat2016} are cool relatives 
to the RCB stars \citep{GarciaHernandez2023}, the excitation
of the comparatively longer-period pulsations up to several hundred days
can still be fitted in the picture presented here (cf. discussion of 
episode~D in Sect.~3). The problem with modeling is rather how to 
preserve the pulsations of such stars from catastrophic amplitude 
growth over long times. The differences of the episodic 
fadings episodes of RCB and DY~Per variables will be intriguing challenges 
to understand. For both groups of variable stars, the answer lies probably 
beyond the photospheres.

\bigskip

\textsc{Acknowledgment } NASA's Astrophysics Data System was as 
indispensable as ever. The model-star sequences discussed in this note
could not have been computed without
the \MESA\,~soft\-ware-instrument~--~used here in version
r$21.12.1$~--~\citep{Paxton2019} and Josiah Schwab's service to add
the \AE SOPUS low-temperature opacities to the test-suite case for R
CrB stars.  Postprocessing \MESA{\tt star}\,models and plotting
benefited from Warrick
Ball's \href{https://github.com/warrickball/tomso}{\texttt{tomso}}
module, \texttt{numpy }\citep{harris2020array},
\texttt{scipy }\citep{2020SciPy}, and \texttt{matplotlib }\citep{Hunter2007},
respectively.

\bibliographystyle{aa}
\bibliography{/home/alfred/StarDude/Librarium/StarBase}

\begin{thebibliography}{26}
\expandafter\ifx\csname natexlab\endcsname\relax\def\natexlab#1{#1}\fi

\bibitem[{Clayton(1996)}]{Clayton1996}
Clayton, G.~C. 1996, PASP, 108, 225

\bibitem[{Cox(1980)}]{jpc80}
Cox, J. 1980, {Theory of Stellar Pulsation} (Princeton, New Jersey: Princeton
  University Press)

\bibitem[{Cox {et~al.}(1980)Cox, Wheeler, Hansen, King, Cox, \&
  Hodson}]{Cox1980}
Cox, J.~P., Wheeler, J.~C., Hansen, C.~J., {et~al.} 1980, Space Sci. Rev., 27,
  529

\bibitem[{Crawford {et~al.}(2020)Crawford, Clayton, Munson, Chatzopoulos, \&
  Frank}]{Crawford2020}
Crawford, C.~L., Clayton, G.~C., Munson, B., Chatzopoulos, E., \& Frank, J.
  2020, MNRAS, 498, 2912

\bibitem[{Crawford {et~al.}(2023)Crawford, Tisserand, Clayton, Soon, Bessell,
  Wood, García-Hernández, Ruiter, \& Seitenzahl}]{Crawford2023}
Crawford, C.~L., Tisserand, P., Clayton, G.~C., {et~al.} 2023, MNRAS, 521, 1674

\bibitem[{García-Hernández {et~al.}(2023)García-Hernández, Rao, Lambert,
  Eriksson, Reddy, \& Masseron}]{GarciaHernandez2023}
García-Hernández, D.~A., Rao, N.~K., Lambert, D.~L., {et~al.} 2023, ApJ, 948,
  15

\bibitem[{Gautschy(2023)}]{Gautschy2023}
Gautschy, A. 2023, arXiv:2303.11374 [astro-ph]

\bibitem[{Harris {et~al.}(2020)Harris, Millman, van~der Walt, Gommers,
  Virtanen, Cournapeau, Wieser, Taylor, Berg, Smith, Kern, Picus, Hoyer, van
  Kerkwijk, Brett, Haldane, del R{\'{i}}o, Wiebe, Peterson,
  G{\'{e}}rard-Marchant, Sheppard, Reddy, Weckesser, Abbasi, Gohlke, \&
  Oliphant}]{harris2020array}
Harris, C.~R., Millman, K.~J., van~der Walt, S.~J., {et~al.} 2020, Nature, 585,
  357

\bibitem[{Hunter(2007)}]{Hunter2007}
Hunter, J. 2007, CSE, 9, 90

\bibitem[{Lauer {et~al.}(2019)Lauer, Chatzopoulos, Clayton, Frank, \&
  Marcello}]{Lauer2019}
Lauer, A., Chatzopoulos, E., Clayton, G.~C., Frank, J., \& Marcello, D.~C.
  2019, MNRAS, 488, 438

\bibitem[{Lawson \& Cottrell(1997)}]{Lawson1997}
Lawson, W.~A. \& Cottrell, P.~L. 1997, MNRAS, 285, 266

\bibitem[{Nikzat \& Catelan(2016)}]{Nikzat2016}
Nikzat, F. \& Catelan, M. 2016, IBVS, 6190, 1

\bibitem[{Paxton {et~al.}(2011)Paxton, Bildsten, Dotter, Herwig, Lesaffre, \&
  Timmes}]{Paxton2011}
Paxton, B., Bildsten, L., Dotter, A., {et~al.} 2011, ApJS, 192, 3

\bibitem[{Paxton {et~al.}(2013)Paxton, Cantiello, Arras, Bildsten, Brown,
  Dotter, Mankovich, Montgomery, Stello, Timmes, \& Townsend}]{Paxton2013}
Paxton, B., Cantiello, M., Arras, P., {et~al.} 2013, ApJS, 208, 4

\bibitem[{Paxton {et~al.}(2015)Paxton, Marchant, Schwab, Bauer, Bildsten,
  Cantiello, Dessart, Farmer, Hu, Langer, Townsend, Townsley, \&
  Timmes}]{Paxton2015}
Paxton, B., Marchant, P., Schwab, J., {et~al.} 2015, ApJS, 220, 15

\bibitem[{{Paxton} {et~al.}(2019){Paxton}, {Smolec}, {Schwab}, {Gautschy},
  {Bildsten}, {Cantiello}, {Dotter}, {Farmer}, {Goldberg}, {Jermyn}, {Kanbur},
  {Marchant}, {Thoul}, {Townsend}, {Wolf}, {Zhang}, \& {Timmes}}]{Paxton2019}
{Paxton}, B., {Smolec}, R., {Schwab}, J., {et~al.} 2019, ApJS, 243, 10

\bibitem[{Saio(2008)}]{Saio2008}
Saio, H. 2008, in ASP Conference Series, Vol. 391, Hydrogen-deficient stars, 69

\bibitem[{Saio {et~al.}(1998)Saio, Baker, \& Gautschy}]{SBG98}
Saio, H., Baker, N., \& Gautschy, A. 1998, MNRAS, 294, 622

\bibitem[{Schwab(2019)}]{Schwab2019}
Schwab, J. 2019, ApJ, 885, 27

\bibitem[{Soszyński {et~al.}(2009)Soszyński, Udalski, Szymański, Kubiak,
  Pietrzyński, Wyrzykowski, Szewczyk, Ulaczyk, \& Poleski}]{Soszynski2009}
Soszyński, I., Udalski, A., Szymański, M.~K., {et~al.} 2009, AcA, 59, 335

\bibitem[{Tisserand {et~al.}(2020)Tisserand, Clayton, Bessell, Welch, Kamath,
  Wood, Wils, Wyrzykowski, Mróz, \& Udalski}]{Tisserand2020}
Tisserand, P., Clayton, G.~C., Bessell, M.~S., {et~al.} 2020, A\&A, 635, A14

\bibitem[{Udalski(2008)}]{Udalski2008}
Udalski, A. 2008, AcA, 58, 187

\bibitem[{Virtanen {et~al.}(2020)Virtanen, Gommers, Oliphant, Haberland, Reddy,
  Cournapeau, Burovski, Peterson, Weckesser, Bright, {van der Walt}, Brett,
  Wilson, Millman, Mayorov, Nelson, Jones, Kern, Larson, Carey, Polat, Feng,
  Moore, {VanderPlas}, Laxalde, Perktold, Cimrman, Henriksen, Quintero, Harris,
  Archibald, Ribeiro, Pedregosa, {van Mulbregt}, \& {SciPy 1.0
  Contributors}}]{2020SciPy}
Virtanen, P., Gommers, R., Oliphant, T.~E., {et~al.} 2020, Nature Methods, 17,
  261

\bibitem[{Wong \& Bildsten(2023)}]{Wong2023}
Wong, T. L.~S. \& Bildsten, L. 2023, arXiv:2311.10158 [astro-ph]

\bibitem[{Wood(1976)}]{Wood1976}
Wood, P.~R. 1976, MNRAS, 174, 531

\bibitem[{Yoon \& Cantiello(2010)}]{Yoon2010}
Yoon, S.-C. \& Cantiello, M. 2010, ApJ, 717, L62

\end{thebibliography}

\end{document}